\documentclass{iopart}
\usepackage{iopams}
\usepackage{bm}
\usepackage{cite}
\usepackage{graphicx}

\begin{document}

\title{Optimizing omnidirectional reflection
by multilayer mirrors}

\author{T Yonte$^\dag$, J J Monz\'on$^\dag$,
A. Felipe$^\ddag$ and L L S\'{a}nchez-Soto$^\dag$}

\address{$^\dag$ Departamento de \'Optica,
Facultad de F\'{\i}sica, Universidad Complutense,
28040~Madrid, Spain}

\address{$^\ddag$ Departamento de Estad\'{\i}stica
e Investigaci\'on Operativa I,
Facultad de Matem\'aticas, Universidad Complutense,
28040 Madrid, Spain}

\begin{abstract}
Periodic layered media can reflect strongly
for all incident angles and polarizations
in a given frequency range. Quarter-wave
stacks at normal incidence are commonplace
in the design of such omnidirectional
reflectors. We discuss alternative design
criteria to optimize these systems.

\end{abstract}

\section{Introduction}

At optical frequencies,  metallic mirrors reflect
strongly for any angle of incidence and any
polarization. However, they display dissipative
losses, which constitutes a drawback in practical
applications. Photonic crystals~\cite{dow} were
originally proposed by Yablonovitch~\cite{Yab87}
to solve this problem: they are periodically
microstructured materials that reflect in stop
(forbidden) bands within which light propagation
is not possible in an infinite structure. Since
they are made from transparent materials,
photonic crystals can be almost free of losses
at any prescribed frequency.

In the one-dimensional case, a photonic crystal
is nothing other than a periodic multilayer.
Although much attention has been payed to
dielectric Bragg mirrors consisting of
alternating low- and high-index layers,
certain aspects of the reflection by periodic
layered media are universal: for $N$ periods
the reflectance goes to unity as $N¨^{-2}$
at the band edges, while tends to unity
exponentially at the band gaps~\cite{Lek87,Yeh88}.

This means that, in practice, not very many
periods are needed to have a stop band. One
is then led to consider stacks of $N$ periods
(which are often called finite periodic
structures) and apply to them conditions that
are valid only when the system is strictly
infinite. Recently~\cite{Mon03,Bar03},
we have put forward this problem and provided
an alternative framework for dealing with
these finite periodic structures: the
trace of the basic period allows us to
classify them into three classes with quite
different properties.

Concerning the performance of these structures, it
is indisputable that Bragg quarter-wave stacks
(designed for normal incidence) are the most
thoroughly studied in connection with omnidirectional
reflection~\cite{Fin98,Dow98,Yab98,Chi99,Sou99,Lek00}.
In spite of this, the current interest in extreme
ultraviolet~\cite{euv} and soft x-ray~\cite{x} optics
is driving a great deal of work on new methods for
optimizing the design of multilayer mirrors. In
addition to the simple but cumbersome optimization
by eye, only recently more sophisticated techniques
have been started to be used~\cite{San01}: relevant
examples include the downhill simplex
algorithm~\cite{Joe95}, the systematic search in
the parameters space~\cite{Mao99}, the simulating
annealing~\cite{Pow01}, the needle variation
technique~\cite{Pro99}, the Levenberg-Marquardt
algorithm~\cite{Win98}, or genetic algorithms~\cite{Mar95}.
The aim of this paper is to provide an alternative
optimization criterion that has the virtue of using
a simple analytical figure of merit with a very
clear physical meaning.

\section{Notations and general relations}

We start by examining the properties of the
basic period of our structure, which is assumed
to be lossless. The field amplitudes at each
side (called ambient and substrate) of the unit
cell are related by the $2 \times 2$ complex
transfer matrix $\mathsf{M}_{as}$ that can be
expressed as~\cite{Mon99a,Mon99b}
\begin{equation}
\label{Mlossless}
\mathsf{M}_{as} =
\left (
\begin{array}{cc}
1/T_{as} & R _{as}^\ast/T_{as}^\ast \\
R_{as}/T_{as} & 1/T_{as}^\ast
\end{array}
\right )
\equiv
\left (
\begin{array}{cc}
\alpha & \beta \\
 \beta^\ast & \alpha^\ast
\end{array}
\right )  ,
\end{equation}
where $R_{as}$ and $T_{as}$ are, respectively,
the overall reflection and transmission
coefficients for a wave incident
from the ambient. Note that for identical
ambient and substrate media we have
$ \det \mathsf{M}_{as}= +1$, which is
equivalent to $|R_{as}|^2  + |T_{as}|^2 = 1$.

We take as known the theory of reflection
from multilayers~\cite{Yeh88} and its main
result for our purposes, namely that strong
reflection will occur when $[\mathrm{Tr} \
( \mathsf{M}_{as})]^2 > 4$ (these conditions,
one for the $s$ polarization and one for
the $p$ polarization, locate the band stops
for each basic polarization).

When we consider a finite periodic system
that consists of $N$ basic periods, it is
possible to show that in the stop bands
the reflectance takes the general
form~\cite{Yeh88,Mon03,Coj01}
\begin{equation}
\mathfrak{R}^{(N)}  =
\frac{ | \beta|^2 } {| \beta |^2 +
[\sinh (\chi) / \sinh(N \chi ) ]^2} ,
\end{equation}
where $\cosh (\chi) =  \mathrm{Re} (\alpha).$
We are considering only positive values
of $\mathrm{Re} (\alpha)$ since negative values
can be treated much in the same way.

In practice, it is usual that the basic period
of the structure consists of two thin
homogeneous dielectric slabs with low,  $n_\ell$, and
high, $n_h$, indices of refraction and corresponding
thicknesses $d_\ell$ and $d_h$. These Bragg structures
are also appropriately denoted as [LH]$^N$, where $
N$ is the total number of periods. In such a case,
the condition $[\mathrm{Tr} \ (\mathsf{M}_{as})]^2
> 4$, required to have a stop band, can be written
as~\cite{Lek00}
\begin {equation}
\label{bands}
| \mathrm{Re} \ (\alpha) | =
|\cos \delta_\ell \cos \delta_h -
\Lambda_{\ell h} \
\sin \delta_\ell \sin \delta_h| > 1 ,
\end {equation}
where $\delta_i= (2 \pi/\lambda) \Delta_i$
is the phase shift of a wave of wavelength
in vacuum $\lambda$ in traversing the layer
$i$th and $\Delta_i$ is the corresponding
optical path, of value
\begin {equation}
\Delta_i =  n_i d_i \cos \theta_i =
d_i \sqrt{n_i^2 - \sin^2 \theta_0} ,
\end {equation}
$\theta_0$ being the angle of incidence. For
simplicity, we have assumed that the system is
imbedded in air.

The function $\Lambda_{\ell h}$ is
\begin{equation}
\Lambda_{\ell h} = \frac{1 + r_{\ell h}^2}
{1 - r_{\ell h}^2} ,
\end {equation}
where $r_{\ell h}$ is the Fresnel reflection
coefficient for the interface $\ell$-$h$. This
function $\Lambda_{\ell h}$ is frequency independent
but takes different forms for $s$ and $p$ polarizations.
However, one can check that, irrespective of the
angle of incidence, the following relation for both
basic polarizations holds:
\begin{equation}
\label{qps}
\frac{\Lambda_{\ell h}(p)}{\Lambda_{\ell h}(s)} =
\left ( \frac{n_\ell}{n_h} \right )^2 < 1 .
\end {equation}
Due to the restriction (\ref{qps}),
whenever Eq.~(\ref{bands}) is fulfilled
for $p$ polarization, it is always
true also for $s$ polarization. In
consequence, the $p$-polarization stop bands
are more stringent than the corresponding
$s$-polarization ones.

Because of historical reasons~\cite{Bor99},
typical use of dielectric mirrors has been
evaluated at normal incidence, with layers
at a quarter-wavelength thick (at the design
frequency):
\begin{equation}
\label{l4in}
d_\ell = \frac{\lambda_\ell}{4} =
\frac{\lambda}{4 n_\ell} ,
\qquad
d_h = \frac{\lambda_h}{4} =
\frac{\lambda}{4 n_h} .
\end{equation}
The optical paths are equal and
maximum reflection occurs at the
frequency
\begin{equation}
\omega_0 = \frac{\pi}{2} \frac{c}{n_\ell d_\ell}
= \frac{\pi}{2} \frac{c}{n_h d_h} ,
\end{equation}
which is the center of the stop band.

\section{Optimization strategy}

In order to explain the optimization criterion
we wish to propose, in Fig.~1 we have plotted
the reflectance $\mathfrak{R}^{(N)}$ of a
Bragg [LH]$^N$ structure as a function of the
angle of incidence for several values of the
number of periods $N$ and for $p$ polarization.
The layer thicknesses are chosen as in Eq.~(\ref{l4in}).

\begin{figure}[h]
\centering
\resizebox{0.75\columnwidth}{!}{\includegraphics{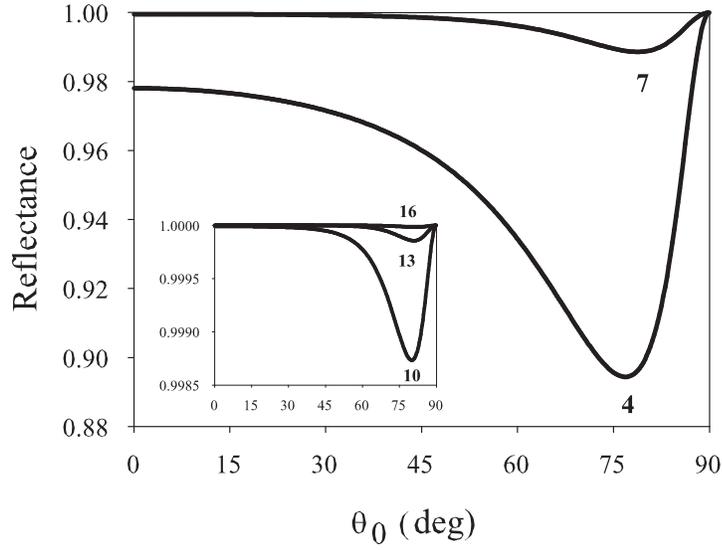}}
\caption{Reflectance for Bragg [LH]$^N$
structures with layers at a quarter-wavelength
thick at $\lambda = 10 \ \mu$m and $p$ polarization
as a function of the angle of incidence. The
refractive indices are $n_\ell = 1.75$ and $n_h =
3.35$. Each curve is labeled with the corresponding
number of periods $N$.}
\end{figure}

It is clear from this figure the well-known fact that
the reflectance tends to unity as $N$ increases.
It is then hardly surprising that the usual
designs found in the literature use
the quarter-wave thicknesses at normal
incidence without raising any doubt about such
assumption. In practice not very many periods
(say $N \sim 10$) are used in the visible, and one
may be tempted to ask whether other thicknesses
could improve the performance of the structure.
Such a problem could be attacked by a straightforward
computation of the reflectance as a function of
layer thicknesses. With the \textit{a priori}
information that the optimum condition is close
to a quarter wave stack, it will not take too
much computational effort to find a reasonable
solution~\cite{Cox00,Doo02,Lee02,Vor02}. However,
our goal here is to provide a more systematic
way of dealing with this question.

A reasonable option for optimizing the system
is that, when $\theta_0$ varies from 0 to
$\pi/2$, the area under the curve
$1 - \mathfrak{R}^{(N)}$ (which is the
transmittance of the system) would be as small
as possible. Therefore, once the materials
and the wavelength are fixed, we propose
\begin{equation}
\label{area}
\mathcal{A} (d_\ell, d_h) = \int_0^{\pi/2}
 [ 1 - \mathfrak{R}^{(N)} (d_\ell, d_h, \theta_0)] \
 d\theta_0 ,
\end{equation}
as figure of merit for the periodic structure.
Alternatively, given the characteristic dip
appearing in the reflectance for $p$ polarization,
one could also impose that this dip would be as
smaller as possible. We have numerically checked
that both criteria give essentially the same results.

\begin{figure}
\centering
\resizebox{0.75\columnwidth}{!}{\includegraphics{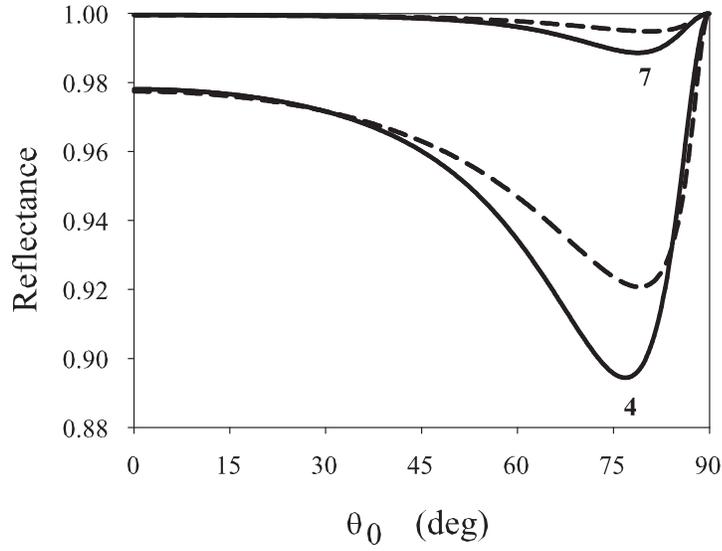}}
\caption{Reflectance for the same Bragg [LH]$^N$ structures (with
$N=4$ and $N=7$) as in Fig.~1. The continuous lines correspond to
quarter-wave thicknesses and the broken lines to the optimum
thicknesses in Table 1.}
\end{figure}

For $s$ polarization there is no dip. The different
behaviour at oblique incidences of $s$ and $p$
reflectances has been analyzed previously~\cite{Lek87}.
Since for a given stack the reflectance $\mathfrak{R}^{(N)}$
is greater for $s$ polarization than for $p$ polarization
for every incidence angle (they are equal only at normal
and grazing incidences), we argue that once the area
$\mathcal{A} (d_\ell, d_h)$ is optimized for $p$
polarization, it is also improved for $s$ polarization,
which seems quite plausible.

We have employed an easy-to-use quasi-Newton
algorithm in order to find the minimum of the
function (\ref{area}), subject only to the
physical conditions $0 < d_\ell/\lambda_\ell \le 1$
and $0 < d_h/\lambda_h \le 1$, because the
periodic character of the solutions. The numerical
results of this optimization are shown in Table 1.
Note that the use of adimensional thicknesses
$d_\ell/\lambda_\ell$ and $d_h/\lambda_h$ simplifies
the presentation of the results, although by no
means the results are universal: they apply only
to this particular system  because we
are not considering dispersion in refractive
indices (in all the paper we take the wavelength
in vacuum $\lambda = 10 \mu$m). In any case, the
proposed optimization strategy is independent
of the particular example under study.

For a better understanding of the performance
of our optimized structure, in Fig.~2 we have
plotted  the reflectance $\mathfrak{R}^{(N)}$
(with $N=4$ and $N=7$) and $p$ polarization
for both the quarter-wave thicknesses as in
Eq.(\ref{l4in}) and the optimum thicknesses
in Table 1. The improvement is remarkable:
in terms of areas, we have 20 \% and
100 \%, respectively.

\begin{figure}[h]
\centering
\resizebox{0.75\columnwidth}{!}{\includegraphics{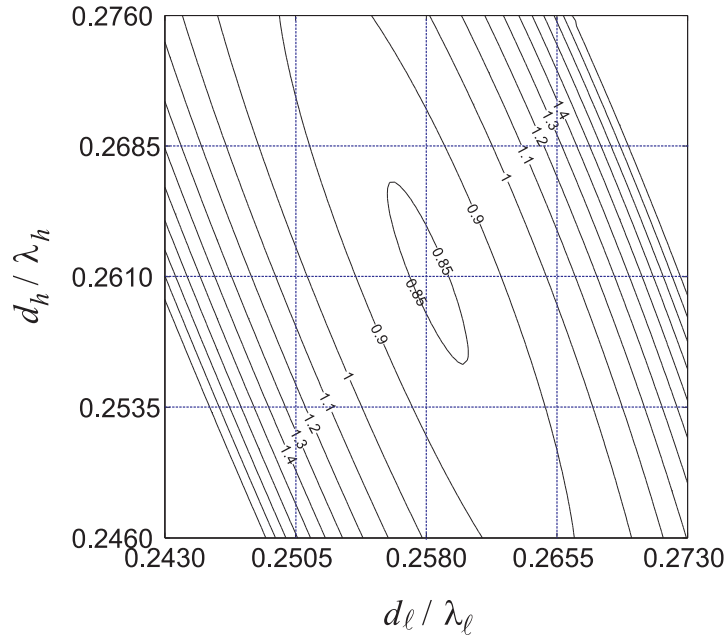}}
\caption{Contour plot of $\mathcal{A}(d_\ell, d_h)$ as a
function of the adimensional thicknesses $d_\ell/\lambda_\ell$
and $d_h/\lambda_h$ for the same structure as in Fig. 2
with $N=7$ periods. The labels in the contour lines show
the value of the area (except for a factor $10^{-2}$). The
optimum working point corresponds to the center of the grid.}
\end{figure}

To test the tolerance of the optimum thicknesses
against small deviations, in Fig.~3 we have a contour
plot of $\mathcal{A}(d_\ell, d_h)$ as a function of the
adimensional thicknesses $d_\ell/\lambda_\ell$ and
$d_h/\lambda_h$ for the same structure as in Fig.~2 with $N=7$
periods. The elliptical contours delimit the range of
thicknesses giving a definite value of the area.
Moreover, if we take the projection of the major axis
of the ellipse on the coordinate axes as a qualitative
measure of the maximum tolerance for a given area, we
conclude from Fig.~3 that one must be more careful in
controlling the thickness $d_\ell$ than $d_h$.

\begin{figure}[h]
\centering
\resizebox{0.75\columnwidth}{!}{\includegraphics{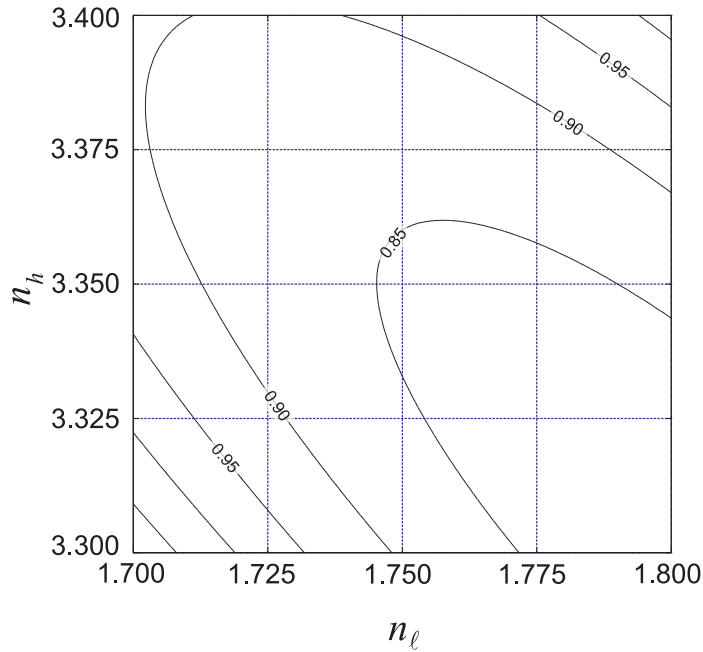}}
\caption{Contour plot of the area as a function of the
refractive indices $n_\ell$ and $n_h$ for the same structure
as in Fig.~3 but with fixed optimum thicknesses $d_\ell$
and $d_h$ given in Table 1. The labels in the contour lines
show the value of the area (except for a factor $10^{-2}$).}
\end{figure}

In the same spirit, one could ask about how critical is
the behaviour of the area under deviations in the values of
refractive indices. In Fig.~4 we have plotted a contour plot
of the area as a function of $n_\ell$ and $n_h$ for the optimum
thicknesses $d_\ell$ and $d_h$ given in Table~1 for $N=7$. We
find again the same kind of elliptical contours as in the
previous figure, but now the variation is much smoother,
indicating that, roughly speaking, the role of refractive
indices is not so crucial as the role of thicknesses.

Obviously, as $N$ grows the improvement in
the area is larger. In fact, table 1 suggests
a considerable improvement that, to some extent,
may be illusory: it only concerns a decrease
in the integrated transmission, while
for the reflection, the improvement is not so
impressive. However, the important point is that
the improvement is just in the dip of the
reflectance. We conclude finally that the method
is especially appropriate for moderate values of
$N$ ($N \sim 10$), which constitute a typical
experimental situation.

The results presented so far hold only for
the given ratio of refractive indices. To show
that the method can be employed for arbitrary
parameters of the multilayers, in Fig.~5 we
have plotted the optimum adimensional thickness
$(d_\ell/\lambda_\ell)_{\mathrm{opt}}$ as a
function of the refractive indices for $N=7$,
while in Fig.~6 we have represented the optimum
values of $(d_h/\lambda_h)_{\mathrm{opt}}$.
We have assigned a zero thickness whenever the
condition $[\mathrm{Tr} \ ( \mathsf{M}_{as})]^2 \ge 4$,
required to have a stop band, is not fulfilled
for some values of the incidence angle $\theta_0$.
The abrupt step is the same in both figures and
gives the boundary of omnidirectional reflection
for the stack.

\begin{figure}
\centering
\resizebox{0.75\columnwidth}{!}{\includegraphics{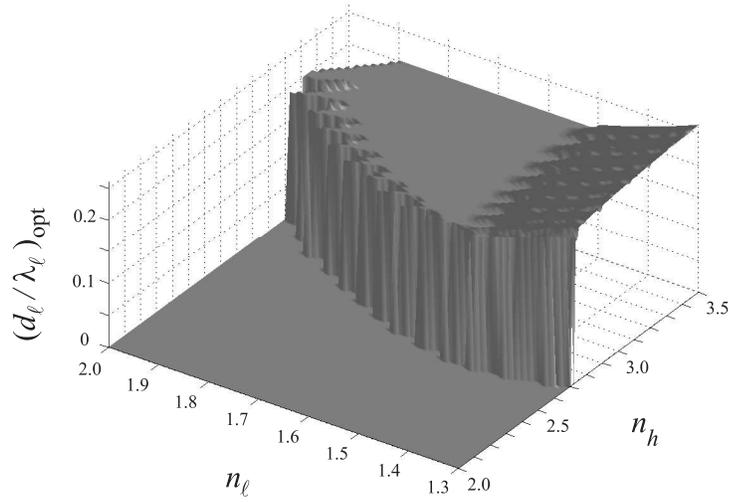}}
\caption{Optimum thickness $(d_\ell/\lambda_\ell)_{\mathrm{opt}}$
as a function of the refractive indices $n_\ell$ and $n_h$ for
a [LH]$^N$ structure with $N=7$, $\lambda = 10 \ \mu$m and $p$
polarization.}
\end{figure}

\begin{figure}
\centering
\resizebox{0.75\columnwidth}{!}{\includegraphics{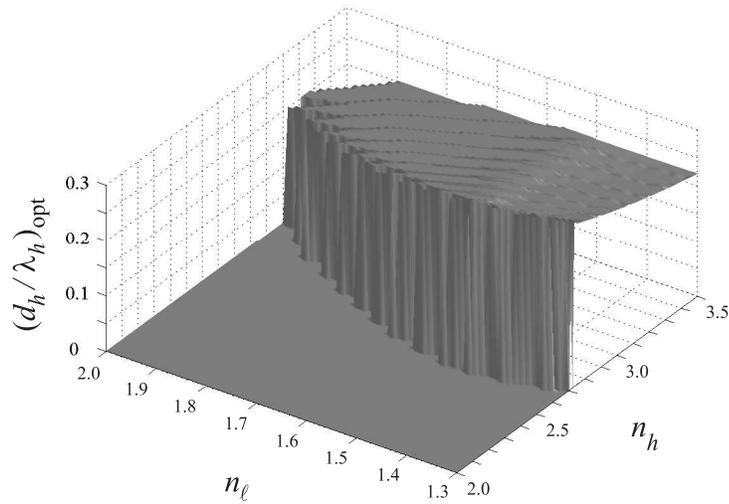}}
\caption{The same as in Fig.~5 but for the optimum thickness
$(d_h/\lambda_h)_{\mathrm{opt}}$.}
\end{figure}

As a general feature, we note that $(d_\ell/\lambda_\ell)_{\mathrm{opt}}
< (d_h/\lambda_h)_{\mathrm{opt}}$ for every pair of allowed
refractive indices. In Fig.~5, the optimum thickness varies
pronouncedly with $n_\ell$ until $n_\ell \simeq 1.5$ and then
it is almost constant, while decreases slowly with $n_h$ in the
region of interest. In Fig.~6 we observe that the optimum thickness
decreases for a fixed value of $n_\ell$ (resp. $n_h$) when
$n_h$ (resp. $n_\ell$) increases, and has a flat variation.
The same qualitative behaviour is also observed
for other values of $N$.

In summary, we have proposed a simple figure of merit that allows
for an improvement in the performance of omnidirectional
reflectors, especially when the number of periods is not too
large. The method can be easily extended to other materials and
wavelengths and predicts optimum thicknesses that can depart
from the usual quarter-wave design.

\begin{table}
\caption{Adimensional thicknesses
$d_\ell/\lambda_\ell$ and
$d_h/\lambda_h$ that optimize the area
for each value of the number of periods $N$.
$\mathcal{A}_\mathrm{opt}$ stands for the
optimum value of the area, while
$\mathcal{A} (1/4)$ represents the value
obtained for quarter-wave design. The
last column shows the percentage of
improvement in these areas computed as
$100 \times [ \mathcal{A} (1/4) - \mathcal{A}_\mathrm{opt}]/
\mathcal{A}_\mathrm{opt}$. The data of the
structure are as in Fig.~1.}

\begin{indented}
\item[]
\begin{tabular}{rccccr}
\br
$N$ & $d_\ell/\lambda_\ell$ & $d_h/\lambda_h$ & $ \mathcal{A}_{\mathrm{opt}}$
& $\mathcal{A}(1/4)$ & $$ Improv. \\
\mr
4 & 0.2567160 & 0.2583615 & 0.6277996 $\times 10^{-1}$ & 0.7539873 $\times 10^{-1}$ & 20 \% \\
5 & 0.2578014 & 0.2596847 & 0.2107099 $\times 10^{-1}$ & 0.2961738 $\times 10^{-1}$ & 41 \%  \\
6 & 0.2582820 & 0.2604825 & 0.7195052 $\times 10^{-2}$ & 0.1199775 $\times 10^{-1}$ & 67 \%  \\
7 & 0.2582544 & 0.2610819 & 0.2508925 $\times 10^{-2}$ & 0.5007563 $\times 10^{-2}$ & 100 \%  \\
8 & 0.2578614 & 0.2616020 & 0.8916341 $\times 10^{-3}$ & 0.2144915 $\times 10^{-2}$ & 141 \% \\
9 & 0.2572066 & 0.2620937 & 0.3219379 $\times 10^{-3}$ & 0.9386422 $\times 10^{-3}$ & 192 \% \\
10 & 0.2563618 & 0.2625808 & 0.1177155 $\times 10^{-3}$ & 0.4179370 $\times 10^{-3}$ & 255 \%   \\
11 & 0.2553779 & 0.2630740 & 0.4347006 $\times 10^{-4}$ & 0.1886905 $\times 10^{-3}$ & 334 \% \\
12 & 0.2542927 & 0.2635773 & 0.1617585 $\times 10^{-4}$ & 0.8614448 $\times 10^{-4}$ & 433 \% \\
13 & 0.2531341 & 0.2640914 & 0.6055056 $\times 10^{-5}$ & 0.3968423 $\times 10^{-4}$ & 555 \%  \\
14 & 0.2519235 & 0.2646149 & 0.2277057 $\times 10^{-5}$ & 0.1841661 $\times 10^{-4}$ & 709 \% \\
\br
\end{tabular}
\end{indented}

\end{table}


\newpage

\begin{thebibliography}{99}


\bibitem{dow}
A complete and up-to-date bibliography
on the subject can be found at the web site
http://home.earthlink.net/\symbol{126}jpdowling/pbgbib.html.

\bibitem{Yab87}
Yablonovitch E
1987
Inhibited spontaneous emission in
solid-state physics and electronics
\textit{ Phys. Rev. Lett.} \textbf{58} 2059-62

\bibitem{Lek87}
Lekner J
1987
\textit{Theory of Reflection}
(Dordrecht: Kluwer)

\bibitem{Yeh88}
Yeh P
1988
\textit{Optical Waves in Layered Media}
(New York: Wiley)

\bibitem{Mon03}
Monz\'{o}n J J, Yonte T and  S\'{a}nchez-Soto L L
2003
Characterizing the reflectance of periodic layered media
\textit{Opt. Commun.} \textbf{218} 43-7

\bibitem{Bar03}
Barriuso A G,  Monz\'{o}n J J and  S\'{a}nchez-Soto L L
2003
General unit-disk representation for periodic multilayers
\textit{Opt. Lett.} \textbf{28} 1501-03

\bibitem{Fin98}
Fink Y,  Winn J N, Fan S, Chen C,
Michel J,  Joannopoulos J D and Thomas E L
1998
A dielectric mmnidirectional reflector
\textit{Science} \textbf{282} 1679-82

\bibitem{Dow98}
Dowling J P
1998
Mirror on the wall: you're omnidirectional after all?
\textit{Science} \textbf{282} 1841-2

\bibitem{Yab98}
Yablonovitch E
1998
Engineered omnidirectional external-reflectivity
spectra from one-dimensional layered interference
filters
\textit{Opt. Lett. }\textbf{23} 1648-9

\bibitem{Chi99}
Chigrin D  N, Lavrinenko A V,
Yarotsky D A and  Gaponenko S V
1999
Observation of total omnidirectional reflection
from a one-dimensional dielectric lattice
\textit{Appl. Phys. A }\textbf{68} 25-8

\bibitem{Sou99}
Southwell W H
1999
Omnidirectional mirror design with quarter-wave
dielectric stacks
\textit{Appl.  Opt.} \textbf{38} 5464-7

\bibitem{Lek00}
Lekner J
2000
Omnidirectional reflection by multilayer dielectric mirrors
\textit{J. Opt. A: Pure Appl. Opt.} \textbf{2} 349-53

\bibitem{euv}
Dobisz E A (ed)
2000
\textit{Emerging Lithographic Technologies}
Proc. SPIE \textbf{3997}

\bibitem{x}
Freund A K, Ishikawa T,  Khounsary A M,  Mancini D C,
 Michette A G, Oestreich S (eds) 2001
\textit{Advances in X-Ray Optics}
Proc. SPIE \textbf{4145}

\bibitem{San01}
S\'anchez del R\'{\i}o M and Pareschi G
2001
Global optimization and reflectivity data fitting
for x-ray multilayer mirrors by means of genetic
algorithms
\textit{Proc. SPIE }\textbf{4145} 88-96

\bibitem{Joe95}
Joensen K D, Voutov P, Szentgyorgyi A,
Roll J, Gorenstein P, Hoghoj P and Christensen F E
1995
Design of grazing-incidence multilayer
supermirrors for hard-x-ray reflectors
\textit{Appl. Opt.} \textbf{34} 7935-44

\bibitem{Mao99}
Mao P H, Harrison F A, Windt D L and Christensen F E,
1995
Optimization of graded multilayer designs
for astronomical X-ray telescopes
\textit{Appl. Opt. }\textbf{38}, 4766-75

\bibitem{Pow01}
Powell K, Tait J M and  Michette A G
2001
Simulated annealing in the design of
broadband multilayers containing more
than two materials
\textit{Proc. SPIE }\textbf{4145}, 254-65

\bibitem{Pro99}
Protopopov V V, Tikhonravov A V, Voronov A V,
Trubetskov M K and DeBell G  K
1999
Optimal design of graded x-ray multilayer
mirrors in the angular and spectral domains
\textit{Proc. SPIE }\textbf{3766}, 320-6

\bibitem{Win98}
Windt D W
1998
IMD-software for modeling the optical
properties of multilayer films
\textit{Comput. in Phys.} \textbf{12} 360-70

\bibitem{Mar95}
Martin S, Rivory J and  Schoenauer M,
1995
Synthesis of optical multilayer systems
using genetic algorithms
\textit{Appl. Opt.} \textbf{34}, 2247-54

\bibitem{Mon99a}
Monz\'{o}n J J and S\'{a}nchez-Soto L L
1999
Lossless multilayers and Lorentz transformations:
more than an analogy
\textit{Opt. Commun.} \textbf{162} 1-6

\bibitem{Mon99b}
Monz\'{o}n J J and  S\'{a}nchez-Soto L L
1999
Fully relativisticlike formulation of
multilayer optics
\textit{J. Opt. Soc. Am. A }\textbf{16}, 2013-18

\bibitem{Coj01}
Cojocaru E
2001
Forbidden gaps in finite periodic and
quasi-periodic Cantor-like dielectric
multilayers at normal incidence
\textit{Appl. Opt.} \textbf{40} 6319-26

\bibitem{Bor99}
Born M and Wolf E
1999
\textit{Principles of Optics}
7 ed. Sec. 1.6.5.
(Cambridge: Cambridge University Press)

\bibitem{Cox00}
Cox S J and  Dobson D C
2000
Band structure optimization of
two-dimensional photonic crystals
in H-polarization
\textit{J. Comput. Phys.} \textbf{158} 214-24

\bibitem{Doo02}
De Dood M J A, Snoeks E, Moroz A and Polman A
2002
Design and optimization of 2D photonic crystal
waveguides based on silicon
\textit{Opt. Quant. Elec.} \textbf{34}, 145-59

\bibitem{Lee02}
Lee H Y, Makino H, Yao T and Tanaka A,
2002
Si-based omnidirectional reflector and transmission
filter optimized at a wavelength of 1.55 $\mu$m.
\textit{Appl. Phys. Lett.} \textbf{81} 4502-04

\bibitem{Vor02}
Vorgul I Y and  Marciniak M  D,
Design and optimisation of multimode
1D photonic band gap waveguide
2002
\textit{Opt. Quant. Electron.}
\textbf{34} 493-503

\end{thebibliography}
\end{document}